\documentclass[%
 reprint,
 amsmath,amssymb,
 aps,
]{revtex4-2}

\usepackage{graphicx}% Include figure files
\usepackage{dcolumn}% Align table columns on decimal point
\usepackage{bm}% bold math
\begin{document}
\title{Multielectron effect in strong-field ionization of aligned nonpolar molecules}

\author{M. Abu-samha}
\affiliation{College of Engineering and Technology, American University of the Middle East, Egaila, Kuwait}%
\author{Lars Bojer Madsen}
\affiliation{Department of Physics and Astronomy, Aarhus University, 8000 Aarhus C, Denmark}

\date{\today}% It is always \today, today,

\begin{abstract}
We revisit strong-field ionization of aligned O$_2$, CO$_2$, and CS$_2$ molecules in light of recent advances in the field of strong-field physics, in particular the inclusion of multielectron polarization in the numerical solution of the time-dependent Schr{\"o}dinger equation (TDSE) within the singe-active-electron approximation. Upon inclusion of multielectron polarization and the associated induced dipole potential in the TDSE model, we obtain angular distributions of total ionization yields which are in better agreement with the available experimental results for CO$_2$ and CS$_2$. For the probed molecules, the main effect of the multielectron polarization on the photoelectron momentum distributions (PMDs) is captured by the introduction of a field at short distances that counteracts the applied external field and leads to a vanishing time-dependent interaction within a certain cut-off radius. Hence, for the PMDs we find no imprints of the long-range part of the laser-induced dipole potential.
\end{abstract}

\maketitle

%\tableofcontents
\section{Introduction}

In strong-field ionization of atomic and molecular targets by intense laser pulses, the  time-dependent Schr{\"o}dinger equation (TDSE) is often used within the single-active-electron (SAE) approximation to describe the ionization process. In this approach, the SAE orbital is propagated in the combined potential of the atom or molecule and the external field. SAE potentials have been successfully applied for some atoms~\cite{TongJPB05,PhysRevLett.81.1207} and molecules~\cite{PhysRevA.81.033416,PhysRevA.81.063406,doi:10.1080/00268970701871007,PhysRevLett.104.223001}. 

While the performance of the SAE approximation is generally accepted, some cases where account of multielectron effects are needed have been discussed in the literature including, for example, the CO~\cite{PhysRevA.96.053421,PhysRevA.101.013433} and OCS~\cite{PhysRevA.102.063111} molecules. For these molecules, the calculated ionization yields and their orientation dependence are affected by multielectron polarization (MEP) or some artifacts of solving the TDSE within the SAE approximation such as the shifting of population from the highest occupied molecular orbital (HOMO) to lower-lying orbitals of the potential. The latter observation was recently illustrated for CO in Ref.~\cite{PhysRevA.101.013433}: the dipole matrix elements which couple the HOMO to lower-lying orbitals are strongly dependent on molecular orientation, and the total ionization yields (TIYs) calculated at different orientation angles were strongly affected by such dipole transitions and the associated artificial shifting of population. As a consequence, for CO, solving the TDSE within the SAE approximation without accounting for MEP predicts the wrong orientation dependence of TIYs~\cite{PhysRevA.101.013433}: the TDSE results predict a maximum TIY at orientation angle $\beta=180^\circ$, that is, when the maximum of the electric ﬁeld points from the O-end to the C-end, which is contrary to experimental observations~\cite{PhysRevLett.108.183001}. The performance of the TDSE method for CO was significantly improved upon inclusion of MEP~\cite{PhysRevA.101.013433}.  Indeed, by extending the TDSE method with a MEP treatment following the approach of Refs.~\cite{PhysRevA.95.023407,0953-4075-51-10-105601}, the inclusion of MEP within the SAE model results in negligible dipole coupling of the HOMO to lower-lying orbitals and the predicted angular distribution of TIYs is in good agreement with  experimental~\cite{PhysRevLett.108.183001} and theoretical data~\cite{PhysRevLett.111.163001,PhysRevA.95.023407}. In this connection, we also note that inclusion of MEP recently led to improved agrement between theory and experiment for odd-even high harmonic generation in CO \cite{PhysRevA.105.023106}.

In the case of CO,  the long-range correction term of the induced dipole potential was not needed to produce the correct trend in TIYs for the CO molecule. It is sufficient to simply turn-off the laser-interaction within a certain molecular cut-off radius, $r_c$. The effective turning-off of the molecule-laser interaction for $r< r_c$  is a consequence of the polarization of the remaining electrons. The electric field associated with this polarization counteracts the applied external field and leads to a cancellation of the interaction. In this sense the  cancellation is a result of the rearrangement of the remaining electrons until they feel no effective field, i.e., a rearrangement  that leads to a field inside $r_c$ that has a direction opposite to the applied field, but with the same magnitude. We will come back to this effect and the value of $r_c$ below.

For OCS, experimental measurements of TIYs and their orientation dependence have attracted much attention in recent literature~\cite{PhysRevA.89.013405,PhysRevA.98.043425,HansenJPhysB2011,Yu_2017,Johansen_2016} due to discrepancies between experimental observations and theoretical predictions. In Ref.~\cite{HansenJPhysB2011}, the experiments were conducted using linearly polarized laser pulses with 30-fs duration at an 800-nm wavelength and intensities of 1.5$\times10^{14}$~W/cm$^2$ and 1.8$\times10^{14}$~W/cm$^2$. At these conditions, the TIY was measured as a function of the orientation angle, and a minimum TIY was obtained at laser polarization parallel to the molecular axis ($\beta=0^\circ$), whereas the maximum TIY was obtained at laser polarization perpendicular to the molecular axis ($\beta=90^\circ$). The orientation dependence of the TIYs presents a challenge to ionization models. For the OCS molecule, the molecular Ammosov-Delone-Krainov model~\cite{PhysRevA.66.033402} and the strong-field approximation (SFA)~\cite{sfa1,sfa2,sfa3} predict the angular dependence of TIYs to follow the orbital structure. In these models, the maximum TIY is predicted at $\beta$ close to 35$^\circ$. Also the Stark-corrected molecular tunneling theory~\cite{PhysRevA.82.053404,Holmegaard2010} and the weak-field asymptotic theory of tunneling, which includes dipole effects~\cite{PhysRevA.84.053423,PhysRevA.87.013406}, did not reproduce the experimental results for the OCS molecule~\cite{PhysRevA.83.023405}. In Ref.~\cite{PhysRevA.102.063111}, we presented theoretical calculations of orientation-dependent TIYs from the HOMO of OCS in a strong linearly polarized laser field by solving the TDSE within the SAE approximation and including MEP effects. The MEP term was represented by an induced dipole term which contains the polarizability of the OCS$^+$ cation parallel to the laser polarization. After adequately accounting for the laser-induced dipole potential associated with the MEP term, the calculated TIYs and their orientation dependence are in good agreement with the experimental measurements reported in Ref.~\cite{HansenJPhysB2011}. The results indicated that the polarizability anisotropy of OCS$^+$ was primarily responsible for the orientation dependence of TIYs from the HOMO of OCS.

\begin{figure}
\includegraphics[width=0.5\textwidth]{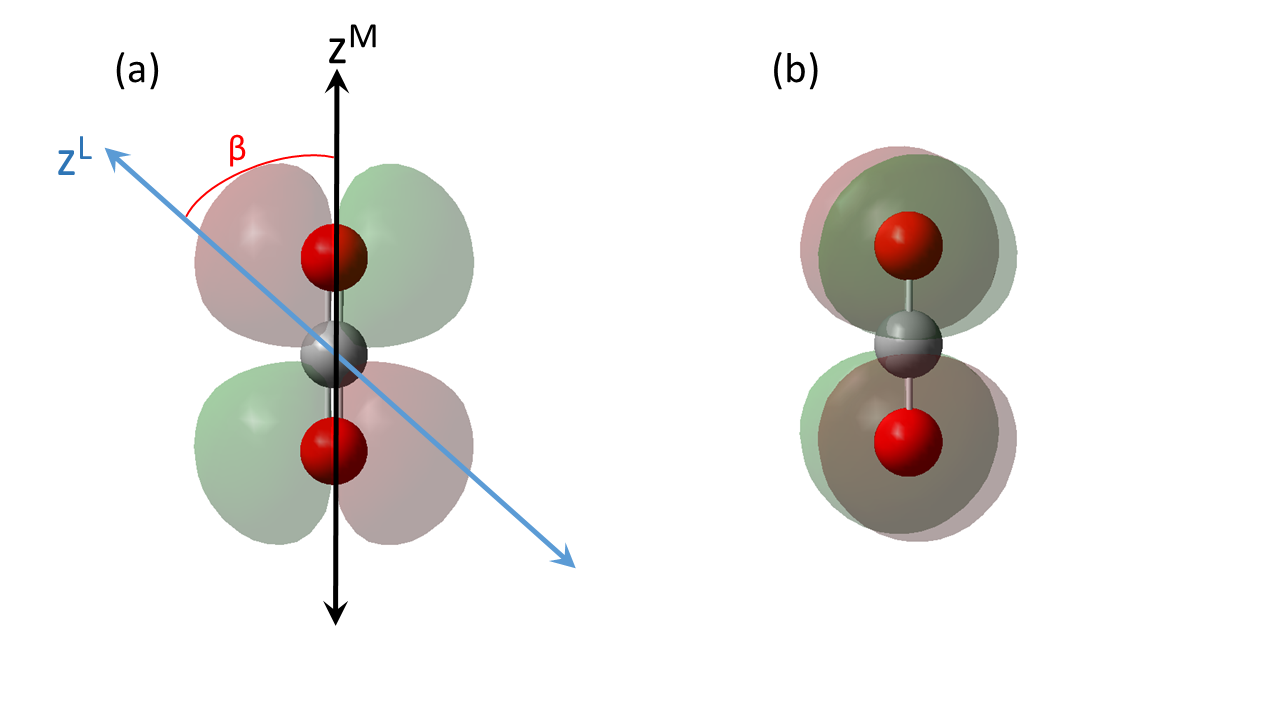}
\caption{\label{co2_homo} Illustration of the degenerate HOMOs of $\pi$ symmetry in CO$_2$ in the molecular fixed (superscript M)  (a) $(xz)^M$ and (b) $(yz)^M$ planes. In (a), the laboratory-frame $z^L$-axis and orientation angle ($\beta$) are illustrated.}
\end{figure}

\begin{table}\caption{\label{molparam} HOMO energies ($E_\text{HOMO}$) and  molecular-fixed frame polarizability components ($\alpha_{xx}$, $\alpha_{zz}$) of the O$_2^+$, CO$_2^+$, and CS$_2^+$ cations as obtained from quantum chemistry calculations at the B3LYP level of theory (MP2 for CO$_2^+$ polarizability)  and  with an aug-cc-pVTZ basis set. All quantities are in atomic units.}
\begin{ruledtabular}
\begin{tabular}{lccc}
&$E_\text{HOMO}$ & $\alpha_{xx}$ & $\alpha_{zz}$ \\
\hline
O$_2^+$  & -0.3199 &  5.08 &  9.41  \\
CO$_2^+$ & -0.3845 &  7.29 & 30.42  \\
CS$_2^+$ & -0.2805 & 28.32 & 76.77  \\
\end{tabular}
\end{ruledtabular}
\end{table}

In this paper, we study strong-field ionization of aligned nonpolar molecular targets: O$_2$, CO$_2$, and CS$_2$.  The HOMOs of these molecules have $\pi_g$ symmetry, see the degenerate HOMOs of CO$_2$ in Fig.~\ref{co2_homo}, and the ionization potentials are comparable; for the ionization potentials we use the negative of the HOMO energies listed in Table~\ref{molparam}. The HOMO($yz$) in Fig.~1(b) has a node in the polarization direction, its contribution to ionization is therefore negligible compared to that of the HOMO($xz$) in Fig.~1(a). Hence only the HOMO($xz$) will be considered in the following. 

For CO$_2$, experimental measurements~\cite{PhysRevLett.98.243001} revealed a narrow angular distribution of TIYs with a maximum TIY at alignment angle $\beta=45^\circ$. For CS$_2$, experimental data~\cite{PhysRevLett.100.093006} are consistent with those for CO$_2$ in that an alignment angle of $\beta=45^\circ$ provides the largest ionization yield. Moreover, the CS$_2$ measurements revealed sharp radial structures in the photoelectron momentum distributions (PMDs) of aligned CS$_2$, an effect that was attributed to Rydberg states brought into resonance by the ac Stark shift~\cite{PhysRevLett.100.093006}.

In our earlier work on CO$_2$~\cite{PhysRevA.80.023401}, strong-field ionization from the HOMO of CO$_2$ was modeled by SAE-TDSE calculations without accounting for MEP, and the theoretical calculations overestimated the yield at alignment angle  $\beta=0^\circ$. Moreover, the TDSE calculations suffered from shifting of population into lower-lying orbitals. Here, we revisit strong-field ionization of CO$_2$ in light of the recent advances in the treatment of MEP within the SAE approximation. The questions we address here are whether the TDSE results for O$_2$, CO$_2$, and CS$_2$ can be improved simply by turning off the external field within $r_c$, thereby minimizing the shifting of population to lower-bound states, or whether the long-range part of the MEP term, represented by a field-induced dipole term, which depends on the polarizability of the cations, see Table~\ref{molparam} and discussion below, has an effect on the angular distributions of TIYs and PMDs. To address these questions, TDSE calculations were conducted using linearly-polarized laser pulses with different laser peak intensities and pulse durations. First, we show how the SAE-TDSE performance is improved once the external field is turned off within $r_c$, we use the CO$_2$ as a proof-of-principle case. Then, we discuss the effects of long-range MEP on alignment dependence of TIYs and PMDs for O$_2$, CO$_2$ and CS$_2$.

The theoretical and computational models are presented in Sec.~\ref{compdet}, followed by results and discussion in Sec.~\ref{res} and conclusions in Sec.~\ref{conc}. Atomic units are used throughout unless otherwise stated.

\section{Theoretical Models}
\label{compdet}
The SAE potentials describing O$_2$, CO$_2$, and CS$_2$ were determined from quantum chemistry calculations~\cite{gamess}, following the procedure detailed in Ref.~\cite{PhysRevA.81.033416}. For O$_2$ and CO$_2$,  the HOMO was obtained in TDSE calculations by propagation in imaginary time. For CS$_2$, the HOMO was obtained following the procedure laid out in Ref.~\cite{PhysRevA.101.013433}.

In our TDSE method, the time-dependent wavefunction $\psi(\vec{r},t)$ of the active orbital is represented by a partial wave expansion in which the spherical harmonics $Y_{lm}(\Omega)$ are used to describe the angular degrees of freedom and a radial grid is used for the time-dependent reduced radial waves, $f_{lm}(r,t)$, i.e.,  
\begin{equation}
    \label{Eq1}
\psi(\vec{r},t)=\sum_{lm} \frac{f_{lm}(r,t)}{r} Y_{lm}(\Omega). 
\end{equation}
The TDSE is solved for an effective one-electron potential describing the interaction with the nucleus, the remaining electrons, and the external field. The TDSE is propagated in the length gauge (LG)~\cite{Kjeldsen2007a} with a combined split-operator~\cite{PhysRevA.38.6000} Crank-Nicolson method. The electric field, $E(t)$, linearly-polarized along the laboratory-frame $z$-axis (from now on the superscript 'L' denoting the laboratory-fixed frame is skipped for brevity), is defined in terms of the vector potential $A(t)$ by
\begin{equation}
    \vec{E}(t) = -\partial_t \vec{A}(t) = -\partial_t \left(\frac{E_0}{\omega}\sin^2(\pi t/\tau)\cos(\omega t+\phi) \right)\hat{z},
    \label{E_field}
\end{equation}
where $E_0$ is the field amplitude, $\omega$ the angular frequency, and $\phi$ the carrier-envelope phase (CEP) for a laser pulse with duration $\tau$. We use a frequency of $\omega=0.057$~a.u. corresponding to a wavelength of 800~nm. The CEP is kept fixed  ($\phi=-\pi/2$). The radial grid contains up to 4096 points and extends to 320 a.u. For the listed molecules, the size of the angular basis set is limited by setting $l_{max}=30$ ($l_{max}=40$ for convergence tests) in Eq.~(\ref{Eq1}). 

The ATI spectra were produced by projecting $\psi(\vec{r},t=\tau)$ at the end of the laser pulse on scattering states of the Coulomb potential in the asymptotic region, and the TIYs were obtained by integrating the above threshold ionization spectra. This approach was recently implemented successfully to obtain TIYs from oriented  CO~\cite{PhysRevA.101.013433} and OCS~\cite{PhysRevA.102.063111}.

\subsection{Extending TDSE methodology with multielectron polarization}
\label{sec:mep}
The theory for the effect of MEP on strong-field ionization was developed in Refs.~\cite{PhysRevLett.95.073001,doi:10.1080/09500340601043413,PhysRevA.82.053404,Holmegaard2010}. For the nonpolar molecules addressed here, in the case when MEP is taken into consideration, the effective potential describing the interaction of the active electron with the core and the time-dependent external field is given asymptotically at large distances as
\begin{equation}
\label{saepot2}
V_{eff}(\vec{r},t) = \vec{r}\cdot\vec{E}(t) -\frac{Z}{r}  - \frac{ \vec{\mu}_{ind}\cdot \vec{r} }{r^3}  \cdots,
\end{equation}
where $\vec{\mu}_{ind}$ is the induced dipole of the cation. Notice that the SAE potential used in most TDSE calculations is missing the induced dipole potential $- \vec{\mu}_{ind}\cdot \vec{r} /r^3$. The MEP term is expressed as $- \vec{\mu}_{ind}\cdot \vec{r} /r^3 = -(\bm{\alpha}\cdot \vec{E}(t) )\cdot\vec{r}/r^3$, where $\bm{\alpha}$ denotes the polarizability tensor. For the linear molecules considered here, and a linearly polarized field with polarization along the $z$ direction,  the product $\bm{\alpha}\cdot\vec{E}(t)$ simplifies to $\left(\alpha_{\perp}E_z,0, \alpha_{||}E_z \right)$. In the laboratory frame, the components perpendicular ($\alpha_{\perp}$) and parallel ($\alpha_{||}$) to the laser polarization can be obtained from the  molecular-fixed frame components, $\alpha_{xx}$ and $\alpha_{zz}$  in Table~\ref{molparam} by a simple rotation: 
\begin{eqnarray}
    \alpha_{\perp} = (\alpha_{zz}-\alpha_{xx})\sin(\beta)\cos(\beta), \\
    \alpha_{||}=\alpha_{xx}\sin^2(\beta) +\alpha_{zz}\cos^2(\beta).
\end{eqnarray}
 At orientation angles $\beta=0^\circ$ or 90$^\circ$,  $\alpha_{\perp}$ and the corresponding induced dipole component vanishes. At other orientation angles, however, there is a non-vanishing contribution to the induced dipole from $\alpha_{\perp}$. We have checked for the molecules considered here that the induced dipole component from $\alpha_{\perp}$ can be omitted: For CO$_2$ (CS$_2$) at $\beta=45^\circ$ within a static external field $E_z=-0.035$~a.u., the laboratory-fixed frame induced dipole components are $\mu_{ind,x}$=0.59 (2.60) and  $\mu_{ind,z}=-1.71$ (-6.15)~Debye, as obtained from quantum chemistry calculations in Gaussian~\cite{g16}. For the O$_2$ molecule, the contribution to the induced dipole from $\alpha_{\perp}$  is negligible because its polarizability components and the 
 polarizability anisotropy $(\alpha_{zz}-\alpha_{xx})$ are very small, see Table~\ref{molparam}. 

Based on the preceding discussion, the MEP term is approximated as $- \vec{\mu}_{ind}\cdot \vec{r} /r^3 \approx -\alpha_{||} \vec{E}(t)\cdot\vec{r}/r^3$, where $\alpha_{||}$ is the static polarizability of the cation parallel to the laser polarization axis. A cutoff radius is chosen close to the core at a radial distance
\begin{equation}
   r_{c}=\alpha_{||}^{1/3},
   \label{r_c}
\end{equation}
such that the MEP cancels the external field at $r\le r_c$~\cite{0953-4075-51-10-105601,PhysRevLett.95.073001,doi:10.1080/09500340601043413}. The interaction term including MEP correction is expressed in the LG at each radial grid point $r_i$ as
\begin{equation}
\label{Eq2}
    V_{LG}^{Ext}(r_i,t)= \begin{cases} \left(1-\frac{\alpha_{||}}{r_i^3}\right)E(t)\sqrt{\frac{2}{3}}r_i \bar{P}_{1}(\zeta), \, \, r > r_c \\
    0, ~r\le r_c \\
    \end{cases}
  \end{equation}
where $E(t)$ is the electric field at time $t$, $\bar{P}_1$ is a normalized Legendre function, and $\zeta=\cos(\theta)$ where $\theta$ is the polar angle of the electron coordinate $\vec{r}$. Based on the above equation, one can see that MEP results in two related effects: the electrons in the cation polarize and set up a field that counteracts the externally applied field at short distances. Hence,  the interaction between the single active electron and the laser field is effectively turned-off at  $r \le r_c$~\cite{PhysRevLett.95.073001,doi:10.1080/09500340601043413}. At long range, that is at $r > r_c$ , the MEP is represented by an induced dipole potential with a  magnitude that depends on the external field strength and the polarizability of the cation.

The polarizability components for the O$_2^+$, CO$_2^+$, and CS$_2^+$ cations were obtained from the NIST computational chemistry database (CCCBDB)~\cite{cccbdb}, see Table~\ref{molparam}. The quantum chemistry calculations of molecular polarizability were conducted at the B3LYP (MP2 for CO$_2^+$) level of theory. 

\section{Results and Discussion}
\label{res}
\subsection{Improving the single-active-electron approximation for CO$_2$ }
\begin{figure}
\includegraphics[width=0.5\textwidth]{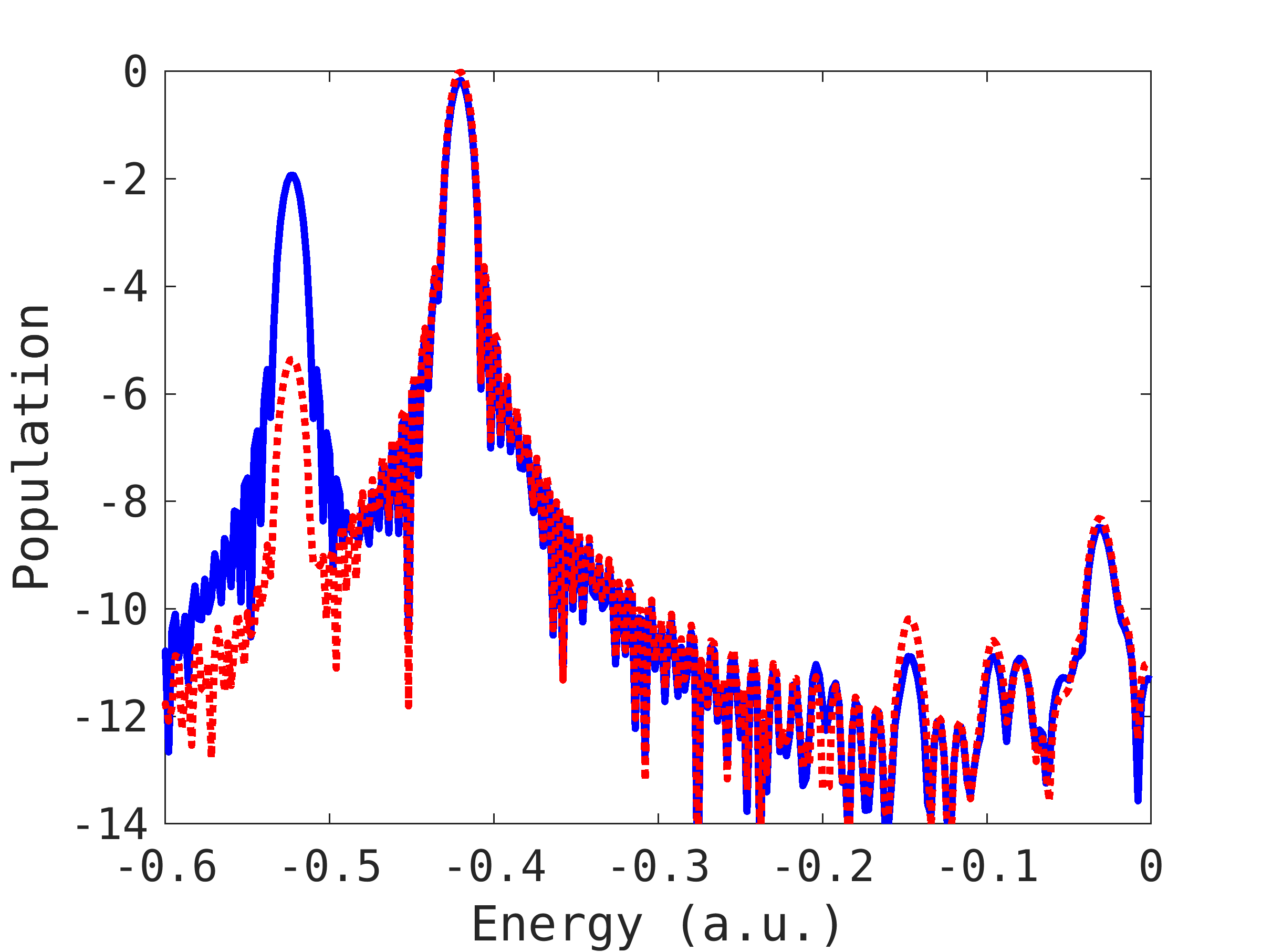}% Here is how to import EPS art
\caption{\label{co2_ati} (Scale is logarithmic on the vertical axis) Bound state spectrum of CO$_2$ after the end of an 800-nm laser pulse containing 5 optical cycles with peak intensity of 5.6$\times$10$^{13}$~W/cm$^2$. The solid (dotted) line denotes TDSE calculations without (with) MEP correction.
}
\end{figure}
Our earlier TDSE calculations for CO~\cite{PhysRevA.82.043413} and CO$_2$~\cite{PhysRevA.80.023401} within the SAE approximation  suffered from shifting of population to lower-lying orbitals of the potential as was recently discussed for CO in Ref.~\cite{PhysRevA.101.013433}. The problem is illustrated for CO$_2$ in Fig.~\ref{co2_ati} (solid line). In this figure, the bound state spectrum is produced for CO$_2$ after probing the molecule by a 5-cycle laser pulse with an intensity of 5.6$\times$10$^{13}$~W/cm$^2$ and $\omega=0.057$~a.u. In Fig.~\ref{co2_ati}, the peaks at  $\sim -0.4$~a.u. and $-0.55$~a.u. correspond to the ($\Pi_g$) HOMO, see Fig.~\ref{co2_homo}, and HOMO-1 of CO$_2$. In the present approach, we consider only the HOMO as being active and it should not  populate the HOMO-1, since that latter orbital is already occupied. 

Here, we consider an improved theory for calculating the TIYs from aligned CO$_2$ which takes into account MEP effect and, thereby, resulting in minimized shifting of population from the HOMO to the lower-lying occupied orbitals of the potential. This is accomplished by turning off the external field within $r_c$ of Eq.~(\ref{r_c}), such that the MEP cancels the external field at $r\le r_c$~\cite{0953-4075-51-10-105601,PhysRevLett.95.073001,doi:10.1080/09500340601043413}.  The improvement of the SAE approximation following the implementation of this effect is illustrated in Fig.~\ref{co2_ati} (dotted line). Here it can be seen that the population in the HOMO-1 (at energy -0.55~a.u.) is reduced by 3-4 orders of magnitude. Meanwhile, the population in the excited states is less sensitive to turning the external field off within the cutoff radius $r_c$. This is promising in particular for cases where strong-field ionization is affected by the excited state manifold. The preceding results and discussion show that it is necessary to turn off the external field within $r_c$ in all TDSE calculations within the SAE approximation. 

\subsection{Effect of long-range multielectron polarization on alignment-dependent total ionization yields and photoelectron momentum distributions of CO$_2$ }

\begin{figure*}
\includegraphics[width=1.0\textwidth]{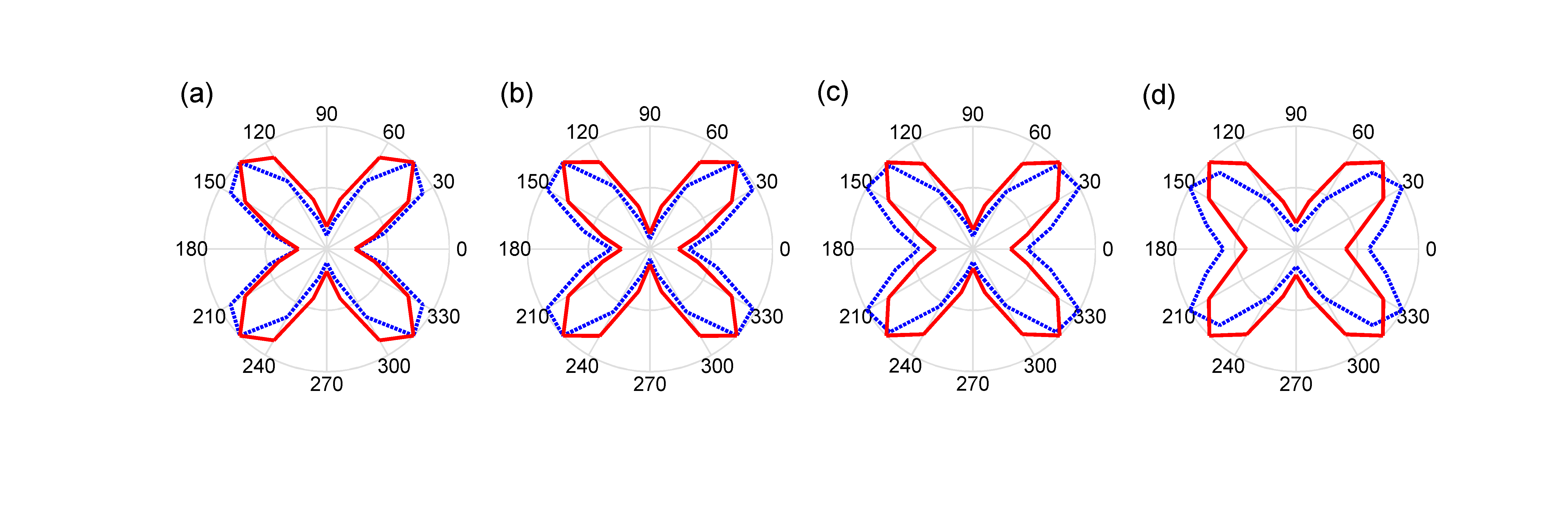}% Here is how to import EPS art
\caption{\label{co2_tiy} Effect of MEP on alignment dependence of TIYs from the HOMO($xz$) orbital of the CO$_2$ molecule at peak intensities of (a) 1.4$\times$10$^{13}$ (b) 3.2$\times$10$^{13}$ (c) 5.6$\times$10$^{13}$ and (d) 8.8$\times$10$^{13}$~W/cm$^2$.  The dashed lines denote no field within $r_{c}$ while solid lines denote full MEP including the long-range laser-induced dipole potential.}
\end{figure*}
Now we discuss, in addition to turning-off the time-dependent interaction below $r_c$, the effect of accounting for the long-range MEP correction in TDSE calculations on the calculated TIYs from the HOMO $xz$ orbital of aligned CO$_2$ (shown in Fig.~\ref{co2_homo} (a)).  The probe laser pulses contain five optical cycles, a wavelength of 800~nm, and peak laser intensities of 1.4$\times$10$^{13}$, 3.2$\times$10$^{13}$, 5.6$\times$10$^{13}$, and 8.8$\times$10$^{13}$~W/cm$^2$. The effect of the long-range MEP correction on angular distributions of TIYs is shown in Fig.~\ref{co2_tiy}. The TIYs were calculated by integrating the photoelectron spectra produced by projecting the wave packet at the end of the laser pulse on Coulomb scattering states in the asymptotic region, see Sec.~\ref{compdet} and Refs.~\cite{PhysRevA.94.023414,PhysRevA.102.063111,PhysRevA.101.013433}. 

\begin{figure*}
\includegraphics[width=0.75\textwidth]{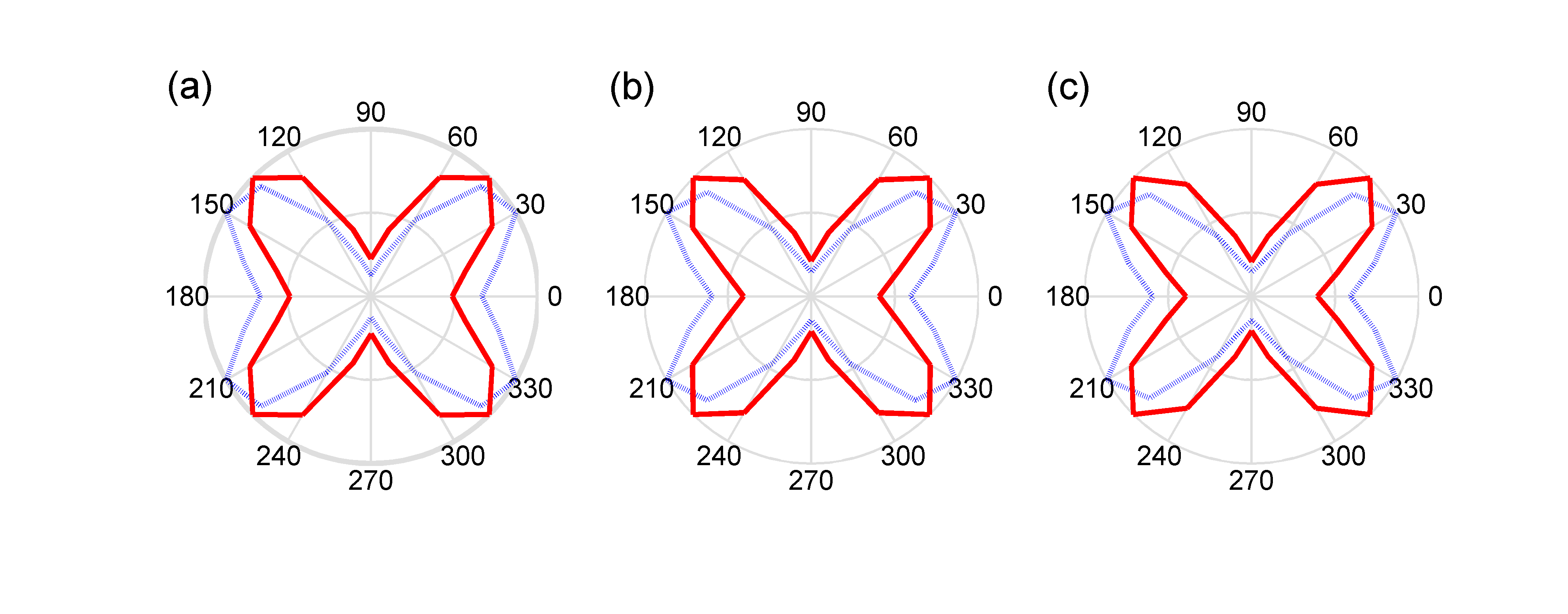}% Here is how to import EPS art
\caption{\label{co2_tiy_ncyc} Effect of MEP on alignment dependence of TIYs from the HOMO($xz$) orbital of the CO$_2$ molecule at peak intensity of  8.8$\times$10$^{13}$W/cm$^2$ for pulses containing (a) 2, (b) 5, and (c) 8 optical cycles. The dotted lines denote no field within $r_{c}$ while solid lines denote full MEP including the long-range laser-induced dipole potential.}
\end{figure*}

We judge the quality of the theoretical TIYs for aligned CO$_2$ based on the following two criteria: the alignment angle of maximum TIY, and the ratio of TIYs at alignment angles $\beta=45^\circ$ and $0^\circ$. Based on early experimental measurements on CO$_2$~\cite{PhysRevLett.98.243001}, the alignment angle of maximum TIY should be $\beta=45^\circ$ and the ratio TIY($\beta=0^\circ$)/TIY($\beta=45^\circ$) is approaching zero. In Fig.~\ref{co2_tiy}, we compare the angular distributions of TIYs for CO$_2$ obtained from TDSE calculations with and without accounting for long-range MEP correction, implemented in Eq.~(\ref{Eq2}). Notice that in both cases, the field is turned off within $r_c$, as we have established based on the results of Sec.~III.A.

From the TDSE calculations for CO$_2$ at the relatively lower intensities of 1.4$\times$10$^{13}$ and 3.2$\times$10$^{13}$~W/cm$^2$, see Fig.~\ref{co2_tiy}~(a) and (b), it is evident  that although the maximum ionization yield is predicted correctly simply by turning off the external field within $r_{c}$, the alignment dependence of the TIY seems to be correctly described only when the full MEP term, including the long-range induced dipole potential, is accounted for. The effect of including long-range MEP seems to decrease the relative ionization probability at  $\beta=30^\circ$. For CO$_2$ at the higher intensities of 5.6$\times$10$^{13}$ and 8.8$\times$10$^{13}$~W/cm$^2$, see Fig.~\ref{co2_tiy}~(c) and (d), the full MEP term, including the long-range induced dipole potential, should be accounted for in order to predict correctly the alignment angle of maximum TIY. Regarding the ratio of TIYs, i.e. TIY($\beta=0^\circ$)/TIY($\beta=45^\circ$), the TDSE model without full MEP treatment (without long-range induced dipole potential ) overestimate the ratio at the higher intensities. 

Next, we discuss the connection between the MEP effect and pulse length, for TDSE calculations at a laser intensity of 8.8$\times$10$^{13}$W/cm$^2$. Alignment-dependent TIYs were calculated for CO$_2$ probed by laser pulses containing 2, 5, and 8 optical cycles, and the results are shown in Fig.~\ref{co2_tiy_ncyc}. For all considered pulse lengths, the orientation angle of maximum TIY shifts from $30^\circ$ (when the long-range MEP term is omitted) to $45^\circ$  (when full MEP is accounted for), indicating that this characteristic of the results is insensitive to pulse length. In short, taking MEP into account improves the prediction of orientation angle ($\beta$) of maximum TIY, in comparison with the experimental measurements, in  particular at high laser intensities. 

\begin{figure*}
\includegraphics[width=0.7\textwidth]{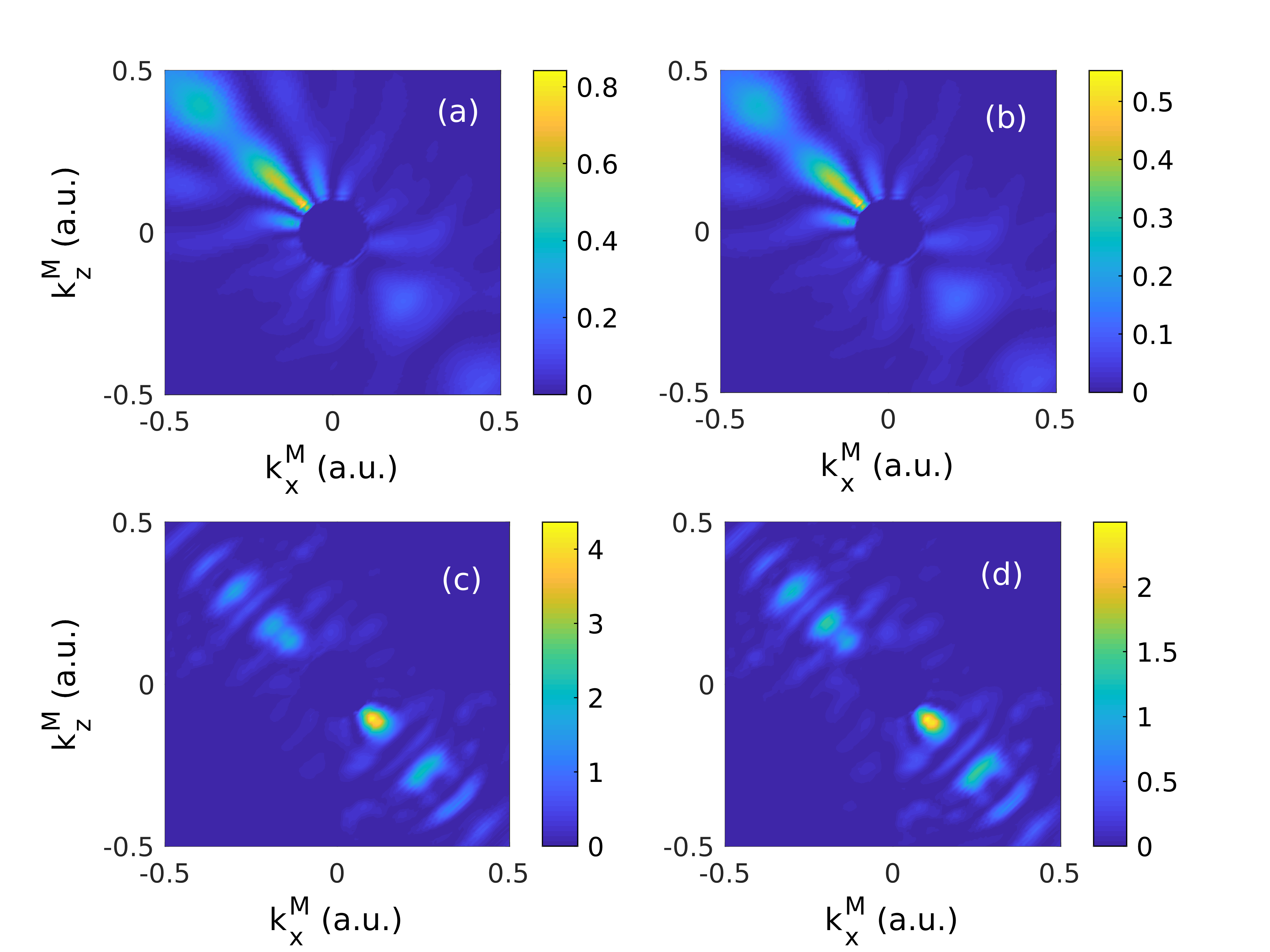}
\caption{\label{co2_pmds} Effect of MEP on laboratory-frame PMDs from the HOMO($xz$) orbital of the CO$_2$ molecule at a peak intensity of 8.8$\times$10$^{13}$W/cm$^2$ and alignment angle $\beta=45^\circ$  for 800 nm pulses containing (a,b) 2 and  (c,d) 8 cycles. In (a,c), the external field is turned off within $r_c$ while in (b,d) the full MEP term is considered.}
\end{figure*}

Finally, we consider the effect of MEP on PMDs for CO$_2$, probed by 800 nm laser pulses containing 2 and 8 optical cycles at a laser intensity of  8.8$\times$10$^{13}$W/cm$^2$. The PMDs were produced by projecting the wave packet on Coulomb scattering states in the asymptotic region, see Sec.~\ref{compdet}. In Fig.~\ref{co2_pmds}, we show PMDs for aligned CO$_2$ at alignment angle of maximum TIY,  $\beta=45^\circ$, obtained with and without long-range MEP correction. From the figure for both 2- and 8-cycle pulses, apart from reduction in overall ionized yield, we do not find characteristic imprints of long-range MEP in the PMDs of CO$_2$. When we compare the 2- and 8-cycle results, we notice the emergence of multiple substructures in the 8-cycle case, which are associated with above-threshold ionization rings, that can be resolved for the longer pulse. 

We have produced PMDs for CO$_2$ at other alignment angles and reached the same conclusion in those cases regarding the effect of the long-range part of the MEP term. In particular we did not find any significant effect in the PMD of the presence or absence of the long-range part of the MEP term at an alignment angle of $\beta=0^\circ$.  We hence face a situation, where the long-range part of the MEP, the induced dipole potential, may change the TIY (Figs. 3 and 4), but not the shape of the PMDs. 

\subsection{Effect of long-range multielectron polarization on alignment-dependent total ionization yields and photoelectron momentum distributions of CS$_2$ }

\begin{figure*}
\includegraphics[width=1.0\textwidth]{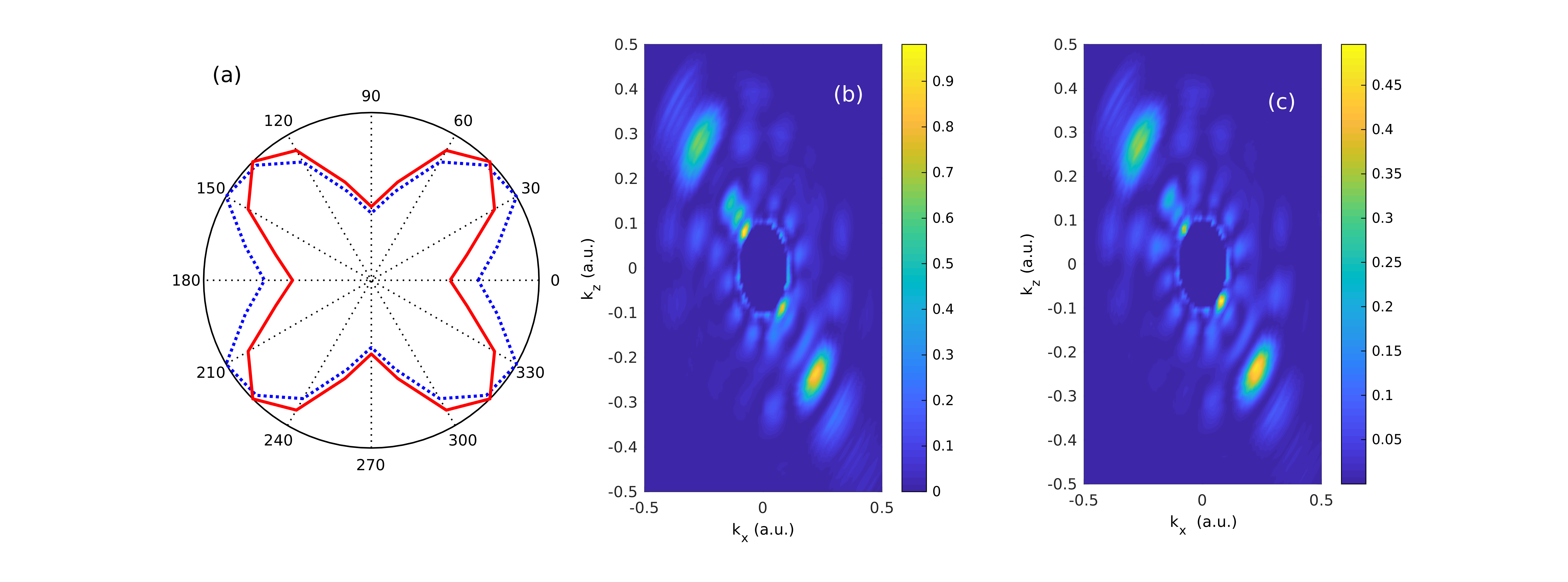}
\caption{\label{cs2_tiy} Effect of MEP on (a) alignment dependence of TIYs and (b,c) photoelectron momentum distributions from the HOMO($xz$) orbital of the CS$_2$ molecule at peak intensity of 4.5$\times$10$^{13}$W/cm$^2$ for pulses containing 5 cycles. In (a), the dotted line denotes no field within $r_{c}$ while the solid line denotes full MEP; the TIYs are given on a relative scale. The PMD in (b) refers to TDSE calculations with no field within $r_{c}$ while (c) refers to full MEP account. }
\end{figure*}

For the CS$_2$ molecule, the HOMO was probed by laser pulses containing 5 optical cycles with a laser peak intensity of 4.5$\times$10$^{13}$W/cm$^2$. The alignment-dependent TIYs are shown in Fig.~\ref{cs2_tiy}(a), where the dotted line denotes no field within $r_{c}$ while the solid line denotes the full MEP treatment, see Sec.~\ref{sec:mep}. From Fig.~\ref{cs2_tiy}~(a), we clearly see that MEP has a significant effect on the angular distribution of TIYs: the alignment angle $\beta$ of maximum TIY shifts from $\beta=30^\circ$ to $45^\circ$ upon accounting for long-range MEP term in the TDSE treatment. We notice that the effect of MEP on the angular distribution of CS$_2$ closely resembles the result for CO$_2$. This makes sense since the molecules have identical orbital symmetry and comparable ionization energies. Also, the Keldysh parameter ($\gamma=1.38$) for CS$_2$ at a laser intensity of 4.5$\times$10$^{13}$W/cm$^2$ is comparable to the one for CO$_2$ at  8.8$\times$10$^{13}$W/cm$^2$, $\gamma=1.2$, suggesting that these molecules are probed within similar ionization regimes.

Turning to the PMDs of aligned CS$_2$, the effect of MEP is illustrated in Fig.~\ref{cs2_tiy} for aligned CS$_2$ at alignment angle $\beta=45^\circ$. In Fig.~\ref{cs2_tiy}~(b) the external field is turned off within $r_c$, whereas in Fig.~\ref{cs2_tiy}~(c) the full MEP term is accounted for. A comparison of the two images reveals no clear observable imprint of MEP on PMDs of aligned CS$_2$.

\subsection{Effect of long-range multielectron polarization on alignment-dependent total ionization yields and photoelectron momentum distributions of O$_2$ }

\begin{figure*}
\includegraphics[width=1.0\textwidth]{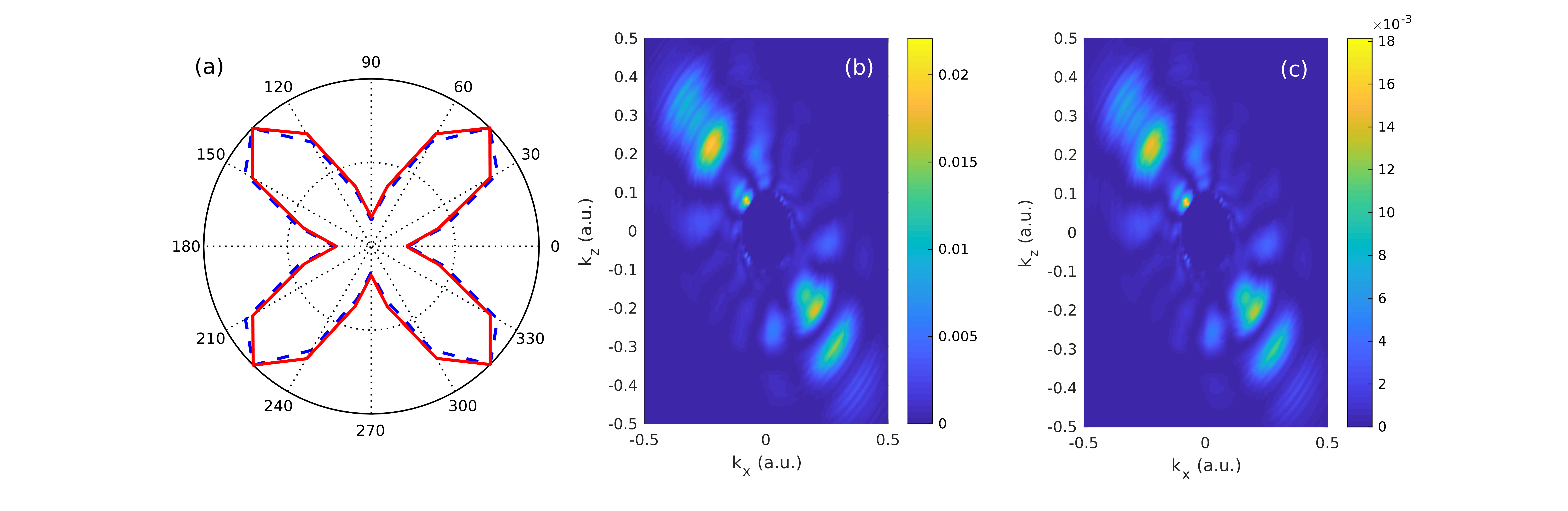}
\caption{\label{o2_tiy} Effect of MEP on (a) alignment dependence of TIYs and (b,c) photoelectron momentum distributions from the HOMO($xz$) orbital of the O$_2$ molecule at peak intensity of 8.8$\times$10$^{13}$W/cm$^2$ for pulses containing 5 cycles. In (a), the dotted line denotes no field within $r_{c}$ while solid line denotes full MEP, and the TIYs are given on a relative scale. The PMD in (b) refers to TDSE calculations with no field within $r_{c}$ while (c) refers to full MEP account. }
\end{figure*}

The HOMO of O$_2$ was probed by laser pulses containing 5 optical cycles with a laser peak intensity of 8.8$\times$10$^{13}$W/cm$^2$. The alignment-dependent TIYs are shown in Fig.~\ref{o2_tiy}(a), where the dotted line denotes no field within $r_{c}$, while the solid line denotes the full MEP treatment, see Sec.~\ref{sec:mep}. From Fig.~\ref{o2_tiy}~(a), we can clearly see that the angular distribution of TIYs has a sharp peak at an alignment angle of $\beta=45^\circ$ and no significant effect of MEP is observed. This can be due to the fact that the O$_2^+$ ion has a relatively small polarizability along the laser polarization ($\alpha_{||}$) compared to CO$_2$ and CS$_2$, see Table~\ref{molparam}.

Now we refer to the effect of MEP on PMDs of O$_2$. We compare the PMDs at alignment angle of maximum TIYs; $\beta=45^\circ$, in Fig.~\ref{o2_tiy}~(b,c). We find no clear observable effect of MEP in the PMDs of O$_2$. This can be understood since the polarizability of the O$_2^+$ cation along the laser polarization ($\alpha_{||}$) is relatively small, compared those for CO$_2$ and CS$_2$. Notice that even in the cases of CO$_2$ and CS$_2$ where the cation polarizability is significantly larger, see Table~\ref{molparam}, we did not find observable effects of MEP in the PMDs.

\section{Conclusions}
\label{conc}
In this work, we revisited strong-field ionization of aligned O$_2$, CO$_2$, and CS$_2$ molecules. We investigated the effect of accounting for MEP in the TDSE methodology within the SAE approximation on alignment dependence of TIYs and PMDs. Generally, MEP results in two related effects. Firstly, the electrons in the cation polarize and set up a field that counteracts the externally applied field at short distances. This effect means that the interaction between the single active electron and the laser field is effectively turned-off at distances smaller than $r_c \sim (\alpha_{||})^\frac{1}{3}$. Secondly, the interaction of the time-dependent laser field with the polarizability of the cation results in an induced dipole potential for $r > r_c$. In general, TDSE calculations within the SAE approximation and without the MEP effects, suffered from shifting of population to lower bound states of the potential, as was demonstrated for CO$_2$. This problem is resolved by turning off the external within a critical radius $r_c$. Now, turning to the effect of long-range MEP correction, taking this correction term into account improves the calculated angular distributions of TIYs for molecules with large polarizabilities of their cations, i.e. CO$_2$ and CS$_2$, in particular at high intensities. The effect of long-range MEP in the PMDs is negligible for the considered molecules, intensities and pulse durations. The present findings are relevant for future investigations of MEP on angular distributions of TIYs and PMDs of oriented polar molecules. 

\section{Acknowledgments}
The numerical results presented in this work were obtained at the Centre for Scientific Computing (CSCAA), Aarhus.

%\section{References}
%\bibliographystyle{unsrt}
%\bibliography{refs}% Produces the bibliography via BibTeX.
\providecommand{\noopsort}[1]{}\providecommand{\singleletter}[1]{#1}%

\end{document}